# Magnon transport in quasi-two-dimensional van der Waals antiferromagnets

Wenyu Xing[1,2], Luyi Qiu[1,2], Xirui Wang[1,2], Yunyan Yao[1,2], Yang Ma[1,2], Ranran Cai[1,2], Shuang Jia[1,2,3], X. C. Xie[1,2,3], and Wei Han[1,2*]

[1] International Center for Quantum Materials, School of Physics, Peking University, Beijing, 100871, China

[2] Collaborative Innovation Center of Quantum Matter, Beijing, 100871, China

[3] CAS Center for Excellence in Topological Quantum Computation, University of Chinese Academy of Sciences, Beijing 100190, China

*Correspondence to: weihan@pku.edu.cn (W.H.)

**The recent emergence of 2D van der Waals magnets down to atomic layer thickness provides an exciting platform for exploring quantum magnetism and spintronics applications. The van der Waals nature stabilizes the long-range ferromagnetic order as a result of magnetic anisotropy. Furthermore, giant tunneling magnetoresistance and electrical control of magnetism have been reported. However, the potential of 2D van der Waals magnets for magnonics, magnon-based spintronics, has not been explored yet. Here, we report the experimental observation of long-distance magnon transport in quasi-two-dimensional van der Waals antiferromagnet $MnPS_3$, which demonstrates the 2D magnets as promising material candidates for magnonics. As the 2D $MnPS_3$ thickness decreases, a shorter magnon diffusion length is observed, which could be attributed to the surface-**

**impurity-induced magnon scattering. Our results could pave the way for exploring quantum magnonics phenomena and designing future magnonics devices based on 2D van der Waals magnets.**

## I. INTRODUCTION

The recent emergence of two-dimensional (2D) van der Waals magnets down to atomic-layer thickness has attracted considerable interest and provided an exciting platform for exploring new physical phenomena in low-dimensional magnetism [1-15]. The long-range ferromagnetic order in 2D magnets has been demonstrated in bilayer $Cr_2Ge_2Te_6$ and single layer $CrI_3$ as a result of magnetic anisotropy [1,2,16]. Shortly, the potential of such van der Waals ferromagnets for spintronics applications has been intensively explored. For example, giant tunneling magnetoresistance in bilayer $CrI_3$ has been demonstrated [17-19], which is much higher compared to conventional single-crystalline-MgO barrier based ferromagnetic tunneling junctions [20,21]. The important role of magnon-assistant tunneling through thin $CrBr_3$ barriers has been shown in the graphene/$CrBr_3$/graphene heterostructures [22]. Because of their 2D nature, efficient electrical control of magnetism in 2D ferromagnetic materials has also been explored [3,23-26], which provides an alternative route towards high-temperature ferromagnetic semiconductors [27,28]. Furthermore, room-temperature 2D ferromagnetism in monolayer van der Waals magnet has also been demonstrated in epitaxial films and ionic liquid gated flakes [8,9,25].

Magnonics refers to the magnon-based spintronics, the use of magnon-mediated spin current for information logic and computing applications [29]. One of the major research directions is to search the suitable magnon transport channel materials, which can propagate magnons over a long distance. Recently, the long distance magnon transport has been demonstrated in 3D ferromagnetic and antiferromagnetic insulators, such as YIG, $Cr_2O_3$, and $Fe_2O_3$ [30-32]. Specially, antiferromagnetic $Fe_2O_3$ offers an electrically tunable magnon channel that could be used for magnonics devices that operate in the absence of any magnetic field [32]. Different from 3D magnetic insulators, such 2D van der Waals magnets offer special platforms for very intriguing quantum transport phenomena, including the spin Nernst effect, topological and Weyl magnons that have already been intensively studied theoretically [33-35]. However, the potential of 2D van der Waals magnets for magnon transport has not been experimentally investigated yet.

Here, we report the magnon transport properties in 2D van der Waals antiferromagnet $MnPS_3$. The long-distance magnon propagation over several micrometers in 2D $MnPS_3$ has been demonstrated, which is comparable to the 3D ferromagnetic insulator yttrium iron garnet (YIG). Systematical studies on the spacing dependence of the nonlocal magnon signal reveal a magnon relaxation length of several micrometers. As the temperature decreases, a longer magnon relaxation length is observed, which could be related to longer magnon lifetimes in such a van der Waals antiferromagnet at lower temperatures. As the 2D $MnPS_3$ thickness decreases, a shorter magnon diffusion length is observed, which could be attributed to the surface-impurity-induced magnon scattering. These experimental results have demonstrated 2D van der Waals magnets as a new platform for the magnonics applications [29,36], and could further pave the way for exploring magnon-dependent quantum transport phenomena in 2D van der Waals magnets [33-35,37].

## II. EXPERIMENTAL

Figure 1(a) illustrates the magnon transport in the quasi-2D van der Waals antiferromgnet $MnPS_3$ devices, where the 2D van der Waals $MnPS_3$ flakes are prepared on the ~300 nm $SiO_2$/Si substrates from bulk MnPS$_3$ single crystals using the mechanical exfoliation method [23]. Bulk MnPS$_3$ single crystals are grown using the chemical vapor transport method. Stoichiometric amounts of high-purity manganese, red phosphorus, and sulfur were sealed into an evacuated quartz tube in a temperature gradient from 780 °C (source region) to 730 °C (growth region) for seven days. Fig. 1(b) shows the crystalline and spin structures of the van der Waals antiferromagnet MnPS$_3$. In each layer of the crystal's *ab* plane, the spins of the Mn atoms are antiferromagnetic coupled with their nearest neighbors, while the interlayer exchange coupling between the Mn spins is ferromagnetic [38]. The Néel temperature ($T_N$) of MnPS$_3$ bulk single crystals is ~79 K, obtained from temperature-dependent magnetization measurement (Fig. S1) under the magnetic fields parallel and perpendicular to the crystal's *ab* plane in a Magnetic Properties Measurement System (MPMS; Quantum Design).

The prepared 2D MnPS$_3$ flakes on ~300 nm SiO$_2$/Si substrates are first identified by a Nikon high-resolution optical microscope, and then fabricated for the nonlocal magnon devices using standard electron-beam lithography and lift-off processes. The electrodes are made of 10 nm thick Pt grown in a magneton sputtering system with a base pressure lower than $8\times10^{-7}$ mbar. The width of the Pt electrodes is ~200 nm. Fig. 1(c) shows the opitcal image of a typical magnon device made on 8 nm 2D MnPS$_3$ flake, where the thickness is determined by atomic force microscopy (Fig. S2). Raman studies show that the electron-beam lithography and device fabrication processes are not damaging the MnPS$_3$ flakes (Fig. S3).

The magnon transport in the quasi-2D van der Waals antiferromagnet MnPS$_3$ is measured using the nonlocal geometry via standard low-frequency lock-in technique in a Physical Properties Measurement System (PPMS; Quantum Design). During the nonlocal magnon transport measurement, a current source (Keithley K6221) is used to provide the low frequency AC current ($f$ = 7 Hz) in the range from 10 to 150 μA in the spin injector Pt electrode, and the nonlocal voltages are measured using lock-in amplifiers (Stanford Research SR830). The voltage probes the magnon-dependent chemical potential due to magnon diffusion in the quasi-2D MnPS$_3$ channel. During the measurement, low noise voltage preamplifiers (Stanford Research SR560) are used to enhance the signal-to-noise ratio.

The magnons are generated in quasi-2D van der Waals antiferromagnet MnPS$_3$ under the left Pt electrode (magnon injector), and then diffuse towards the right Pt electrode (magnon detector) which detects the magnon-mediated spin current via inverse spin Hall effect of Pt in the nonlocal geometry [30-32,39,40]. To perform magnon injection, both electrical means via spin Hall effect of Pt and thermal means via thermal spin injection could be utilized [30], which give rise to the first and second harmonic nonlocal voltages probed at the magnon detector ($V_{1\omega}$ and $V_{2\omega}$), respectively. Both means could be used to investigate the magnon transport properties in magnetic insulators, as demonstrated in previous reports [40-43]. In our experiment with the AC injection curent ($I_{in}$ in Fig. 1(c)) in the range from 10 to 150 μA, only the second harmonic nonlocal voltages could be clearly observed, while no obvious first harmoinc voltrages could be detected (Fig. S4). This result could be attributed to higer effciencey magnon generation via themral means than via the electrical means due to spin Hall effect of Pt. Hence, to probe the magnon transport in quasi-2D van der Waals MnPS$_3$, thermal means is utilized to generate the magnons arising from the temperature gradient at the MnPS$_3$-Pt interface via Joule heating [30,31,41,44].

**III. RESULTS AND DISCUSSION**

Figure 2(a) shows the nonlocal magnon transport and measurement geometry in MnPS$_3$, and Fig. 2(b) shows the second harmonic resistance ($R_{2\omega} = \sqrt{2}V_{2\omega}/I_{in}^2$) as a function of the magnetic field angle ($\varphi$) measured on the nonlocal magnon devices of 8 nm (black) and 16 nm (blue) MnPS$_3$ flakes, with spacing (*d*) of 2 μm for both devices measured at *T* = 2 K. During the measurement, the in-plane static magnetic field (*B*) is held at 9 T, which gives rise to a canted magnetic moment ($\vec{m}$) between Mn$_A$ and Mn$_B$ in different magnetic sub-lattices (Fig. 2(a)). The magnon-mediated spin current ($J_m$) carrying the angular momentum parallel to $\vec{m}$ will be converted to the electron-mediated spin current in Pt. Thus, the spin accumulation direction is parallel to $\vec{m}$. When the magnetic field angle is 90 (-90) degrees, the maximum (minimum) voltages will be detected. On the other hand, if the magnetic field angle is 0 degree, as illustrated in Fig. 2(c), zero voltage signal is detected via inverse spin Hall effect since the spin accumulation direction is parallel to the Pt electrode. The 2D MnPS$_3$ magnon devices are rotated from 0 to 360 degrees in an inplane static magnetic field, leading to the $2\pi$-periodic rotation of $\vec{m}$. No clear signals of first harmonic nonlocal voltage could be detected under the AC injection current from 10 to 150 μA, which could be attributed to lower efficiency of magnon generation via the electrical means compared to the thermal means (Fig. S4). This observation is in accordance with previous reports on ferromagnetic insulator YIG and antiferromagnetic insulator Cr$_2$O$_3$ [30,31,44]. Consistent with expectations for the transport of incoherent magnons [30,31], the second harmonic nonlocal voltage scales quadratically with the injection current (Fig. S5).

Since the 2D MnPS$_3$ is insulating (Figs. S6 and S7), the flow of any charge current in the 2D MnPS$_3$ is forbidden, which rules out any nonlocal voltage arising from charge current leakage in the MnPS$_3$ channel. Furthermore, the absence of the $R_{2\omega}$ signal in the control samples fabricated directly on the SiO$_2$/Si substrates confirms that the $R_{2\omega}$ signal can only be obtained on the MnPS$_3$ devices via magnon transport (Fig. S8).

As discussed earlier, since the magnons are injected via thermal means and the diffusive magnons are detected via the inverse spin Hall effect of Pt, $R_{2\omega}$ is expected to be proportional to $\sin(\varphi)$ [30]:

$$R_{2\omega} = R_{NL} \sin(\varphi) \tag{1}$$

where the $R_{NL}$ is the nonlocal spin signal. The red solid lines in Fig. 2(b) are best fitting curves based on the equation (1), from which $R_{NL}$ is determined to be 99 V/A$^2$ for the 8 nm MnPS$_3$ device and 146 V/A$^2$ for the 16 nm MnPS$_3$ device, respectively. Obviously, a larger nonlocal magnon signal ($R_{NL}$) is observed for 16 nm MnPS$_3$ device, which could be attributed to longer magnon relaxation lengths for thicker MnPS$_3$ and be discussed later in detail. The in-plane magnetic field dependence of the nonlocal magnon signals on both devices is shown in Fig. 2(d) ($T$ = 2 K). As in-plane magnetic field increases, the canted magnetic moment increases, giving rise to the enhancement of the second harmonic nonlocal magnon signals [31]. Fig. 2(e) shows the temperature dependence of the nonlocal magnon signals of both devices at $B$ = 9 T. The nonlocal magnon signals are observed when the temperature is lower than ~ 20 K for the 8 nm MnPS$_3$ device and ~ 30 K for the 16 nm MnPS$_3$ device, which could be attributed to lower Néel temperature for thinner MnPS$_3$, a Heisenberg antiferromagnet [45]. These results demonstrate the potential of using a quasi-2D van der Waals antiferromagnet for the magnon transport, which might have advantages compared to ferromagnetic insulator YIG, such as the capability of functioning in the presence of large magnetic fields and the absence of stray fields [46,47].

To investigate the magnon transport properties of quasi-2D van der Waals antiferromagnet MnPS$_3$, the spacing profile of the magnon-dependent chemical potential ($\mu_m$) is systematically studied. Since the MnPS$_3$ thickness is much smaller than the spacing between the two Pt electrodes, magnon transport is expected to follow the one-dimensional drift-diffusion model [30,48]. As illustrated in Fig. 3(a), $\mu_m$ is expected to exponentially decay as the magnon-mediated spin current diffuses away from the magnon injector in the presence of magnon scatterings. At the distance of $d$ away from the magnon injector, $\mu_m$ can be described by the following expression:

$$\mu_m = \frac{\mu_0}{\lambda} \frac{\exp(\frac{d}{\lambda})}{1-\exp(\frac{2d}{\lambda})} \tag{2}$$

where $\mu_0$ is the magnon-dependent chemical potential in MnPS$_3$ under the Pt injector and $\lambda$ is the magnon relaxation length. Quantatively, the decrease of the nonlocal magnon resistances as a function of $d$ can be expressed by [40,41]:

$$R_{NL} = \frac{C}{\lambda} \frac{\exp(\frac{d}{\lambda})}{1-\exp(\frac{2d}{\lambda})} \tag{3}$$

where $C$ is a constant related to the spin-to-charge conversion effciency of Pt, the magnon injection/detection efficiencies, and the spin-mixing conductances at the interface between MnPS$_3$ and Pt.

Fig. 3(b) shows the normalized nonlocal magnon resitance ($R_{NL}^*$) as a funtion of $d$ for the 16 nm MnPS$_3$ device measured at $B$ = 9 T and $T$ = 2 K and 15 K, repectively. $R_{NL}^*$ is used to take into account of the length effect of the magnon detector, and it is calcualted using the following formula $R_{NL}^* = R_{NL} \times \frac{1}{l_{Pt\_D}}$, where $l_{Pt\_D}$ is the length of the magnon detector. During the investigation of the magnon relaxation lengths, the spacings are purposely chosen to be longer than or equal to 2 μm, since there is a magnetic depletion region close to the Pt injector arising from local Joule heating induced spin Seebeck effect (Fig. S9) [42]. The spin signal probed from the local Pt injector is of opposite sign compared to the nonlocal Pt detector which probes the magnon transport across a spacing that is much bigger than the thickness of MnPS$_3$ (Fig. S9).

To quantitatively determine the magnon relaxatton length, the $\log(R_{NL}^*)$ vs. $d$ measured at various temperatures are plotted in Fig. 3(c). It is clear that the experimental results at various temperatures are all in good agreement with the exponential decay of magnon-dependent chemical potential expected theoretically (solid lines in Fig. 3(c)). Based on the best fitting results of the experimental data, the magnon relaxation lengths of the 16 nm MnPS$_3$ are obtained to be 2.8 $\pm$ 0.4 μm at $T$ = 2 K, 2.0 $\pm$ 0.2 μm at $T$ = 10 K, 1.4 $\pm$ 0.1 μm at $T$ = 15 K, and 1.3 $\pm$ 0.1 μm at $T$ = 20 K, respectively. As shown in Fig. 3(d), the magnon relaxation length increases as temperature decreases, which could be attributed to the enhancement of the magnon lifetime time at lower temperatures. This observation is consistent with previous reports of enhancement of magnon lifetimes at low temperatures in antiferromagnets Cu$_2$V$_2$O$_7$ and MnF$_2$ investigated via spin Seebeck effect measurements [49,50].

Next, the magnon transport and relaxation properties of quasi-2D MnPS$_3$ are investigated as a function of the MnPS$_3$ thickness. For this purpose, various magnon devices are fabricated on MnPS$_3$ flakes with different thicknesses and the magnon relaxation length is obtained via the spacing dependence of the nonlocal spin signal. The exponential decay of magnon-dependent chemical potential has been observed on all the MnPS$_3$ devices, which has been further confirmed on the MnPS$_3$ devices with much larger spacings (Fig. S10). Figs. 4(a) and 4(b) show the temperature dependence of the magnon relaxation lengths for the 8 nm MnPS$_3$ device and the 27 nm MnPS$_3$ device, respectively. The temperature dependences of the magnon relaxation lengths for these devices are similar to that measured on the 16 nm MnPS$_3$ device (Fig. 3(d)).

Fig. 4(c) shows the thickness dependence of the magnon relaxation lengths at $T$ = 2, 5, and 10 K, respectively. Clearly, an increase of the magnon relaxation length is observed as the 2D MnPS$_3$ thickness increases and it reaches ~ 4.7 μm for the 40 nm MnPS$_3$ device. These magnon relaxation lengths for quasi-2D antiferromagnet MnPS$_3$ are comparable to the values obtained in 3D ferromagnetic insulator YIG [30], which makes quasi-2D antiferromagnet MnPS$_3$ a promising candidate for magnon spintronics applications. As 2D MnPS$_3$ thickness decreases, the magnon relaxation length decreases, i.e., a short magnon relaxation length of ~ 1.1 μm is obtained on the 8 nm MnPS$_3$ device. These results indicate the presence of the enhanced magnon scattering for thinner MnPS$_3$. One possible reason is the increase of surface-impurity-induced magnon scattering for thinner MnPS$_3$ films, which gives rise to the strong suppression of the magnon lifetime. For devices fabricated on ultrathin MnPS$_3$ flakes, no clear magnon transport signals have been observed yet (see Fig. S11 for the results from two typical devices on 5 nm and 4 nm MnPS$_3$ with $d$ = 2 μm), which could be attributed to the presence of strong surface-impurity-induced magnon scattering and lower Néel tempeature for thinner MnPS$_3$, a Heisenberg antiferromagnet [45].

## IV. CONCLUSIONS

In conclusion, long-distance magnon transport over several micrometers has been demonstrated in the quasi-two-dimensional van der Waals antiferromagnet MnPS$_3$. Systematical studies on the temperature and MnPS$_3$ thickness dependences of the magnon relaxation lengths have been performed. As the 2D MnPS$_3$ thickness decreases, a shorter magnon diffusion length is observed, which could be attributed to the surface-impurity-induced magnon scattering. Our results demonstrate that van der Waals antiferromagnets provide a 2D platform for magnon spintronics and magnon spin computing [29]. Furthermore, these results could pave the way for the future investigation of novel magnon phenomena in van der Waals 2D magnets, including spin Nernst effect, magnon topological properties, quantum magnon Hall effect, etc [33-35,37].

**Acknowledgement**

We acknowledge the financial support from National Basic Research Programs of China (973 program Grants No. 2015CB921104, No. 2018YFA0305601, and No. 2015CB921102), National Natural Science Foundation of China (NSFC Grants No. 11574006, 11774007, No. 11534001, and No. U1832214) and the Strategic Priority Research Program of the Chinese Academy of Sciences (Grant No. XDB28020100).

**Figure 1**

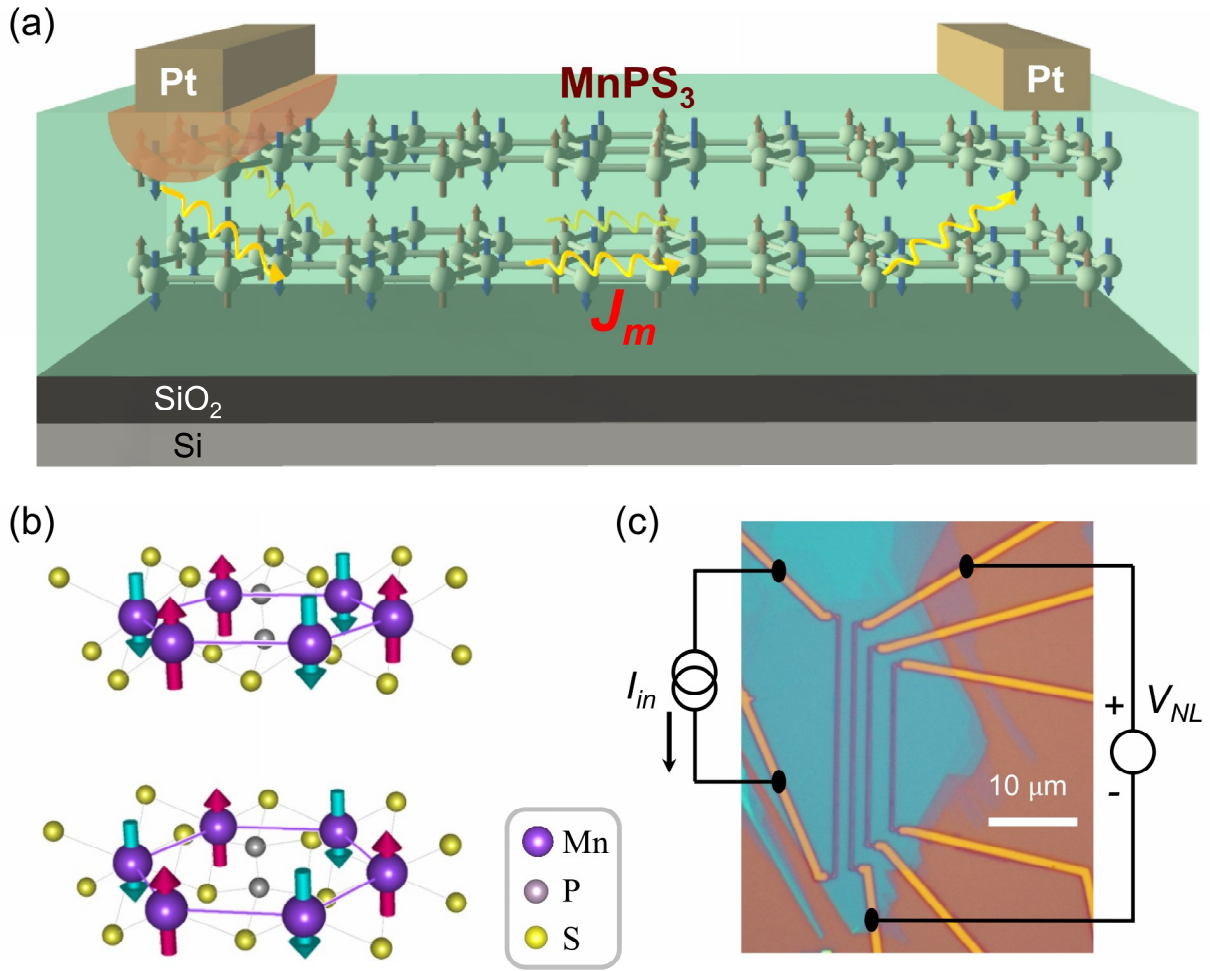

Figure 1 | Illustration of magnon transport in quasi-2D antiferromagnet MnPS$_3$ devices. (a) Schematic of magnon transport in the quasi-2D MnPS$_3$. $J_m$ represents the magnon-mediated spin current. The left Pt electrode is used as magnon injector via thermal spin injection, and the right Pt electrode is used as magnon detector via inverse spin Hall effect. (b) The crystal and spin structures of van der Waals antiferromagnet MnPS$_3$. (c) Schematic of nonlocal measurement on a typical MnPS$_3$ device via low frequency lock-in technique. The thickness of the MnPS$_3$ flake is 8 nm probed by atomic force microscopy (Fig. S2).

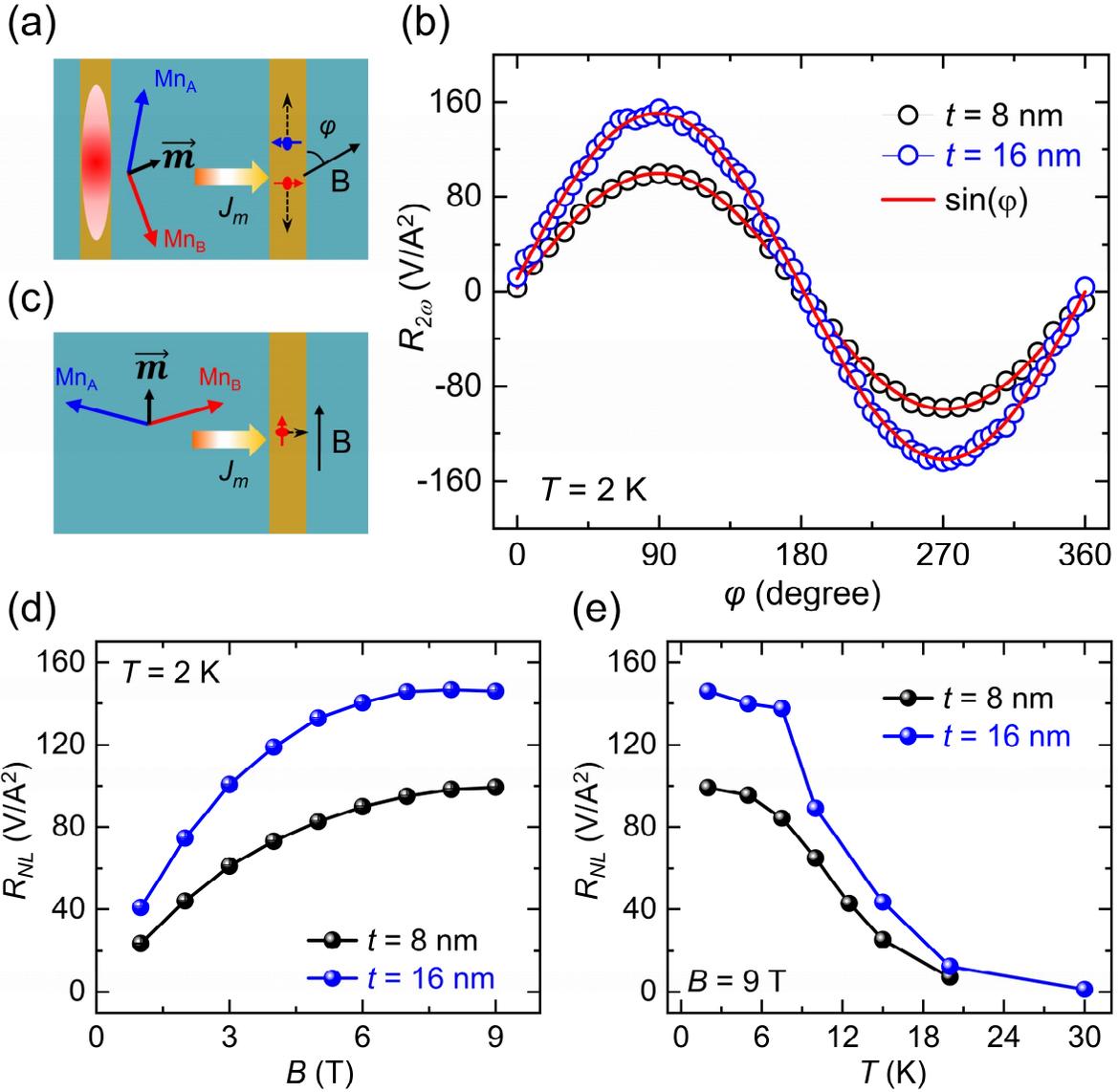

Figure 2 | Nonlocal magnon transport measurements in quasi-2D antiferromagnet MnPS$_3$. (a) Schematic of the generation, transport, and detection of magnon-mediated spin current in MnPS$_3$ magnon devices. (b) The nonlocal resistances of the 8 nm and 16 nm MnPS$_3$ devices ($d$ = 2 μm) as a function of the in-plane magnetic field angle at $T$ = 2 K and $B$ = 9 T. (c) Illustration of the zero voltage signal when the magnetic field angle is 0 degree. (d) The magnetic field dependence of the nonlocal magnon signal for both MnPS$_3$ devices at $T$ = 2 K. (e) The temperature dependence of the nonlocal magnon signal for both MnPS$_3$ devices at $B$ = 9 T.

**Figure 3**

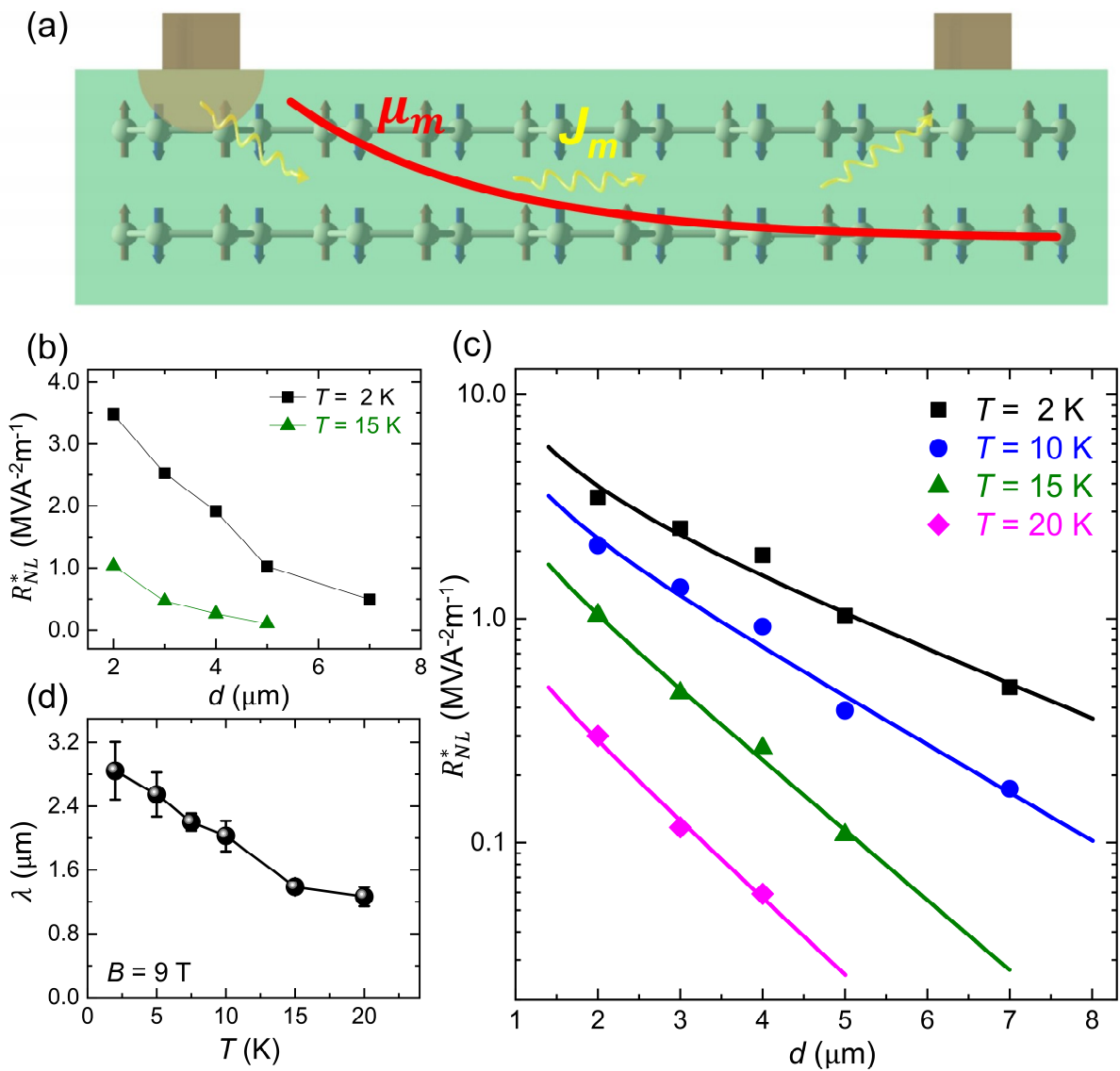

Figure 3 | The measurement of magnonic properties in quasi-2D antiferromagnet MnPS$_3$. (a) Schematic of the exponential decay (solid red line) of magnon-dependent chemical potential ($\mu_m$) during the flow of magnon-mediated spin current ($J_m$). For simplicity, the magnon depletion below the Pt injector is not shown here (see Fig. S9 for details). (b) The spacing (*d*) dependence of the normalized nonlocal spin signals ($R_{NL}^*$) of the 16 nm MnPS$_3$ device measured at *T* = 2 and 15 K. (c) The spacing (*d*) dependence of the normalized nonlocal spin signals ($R_{NL}^*$). The solid lines represent the best fitting results based on equation (3). (d) The temperature dependence of the magnon relaxation length for the 16 nm MnPS$_3$ device.

**Figure 4**

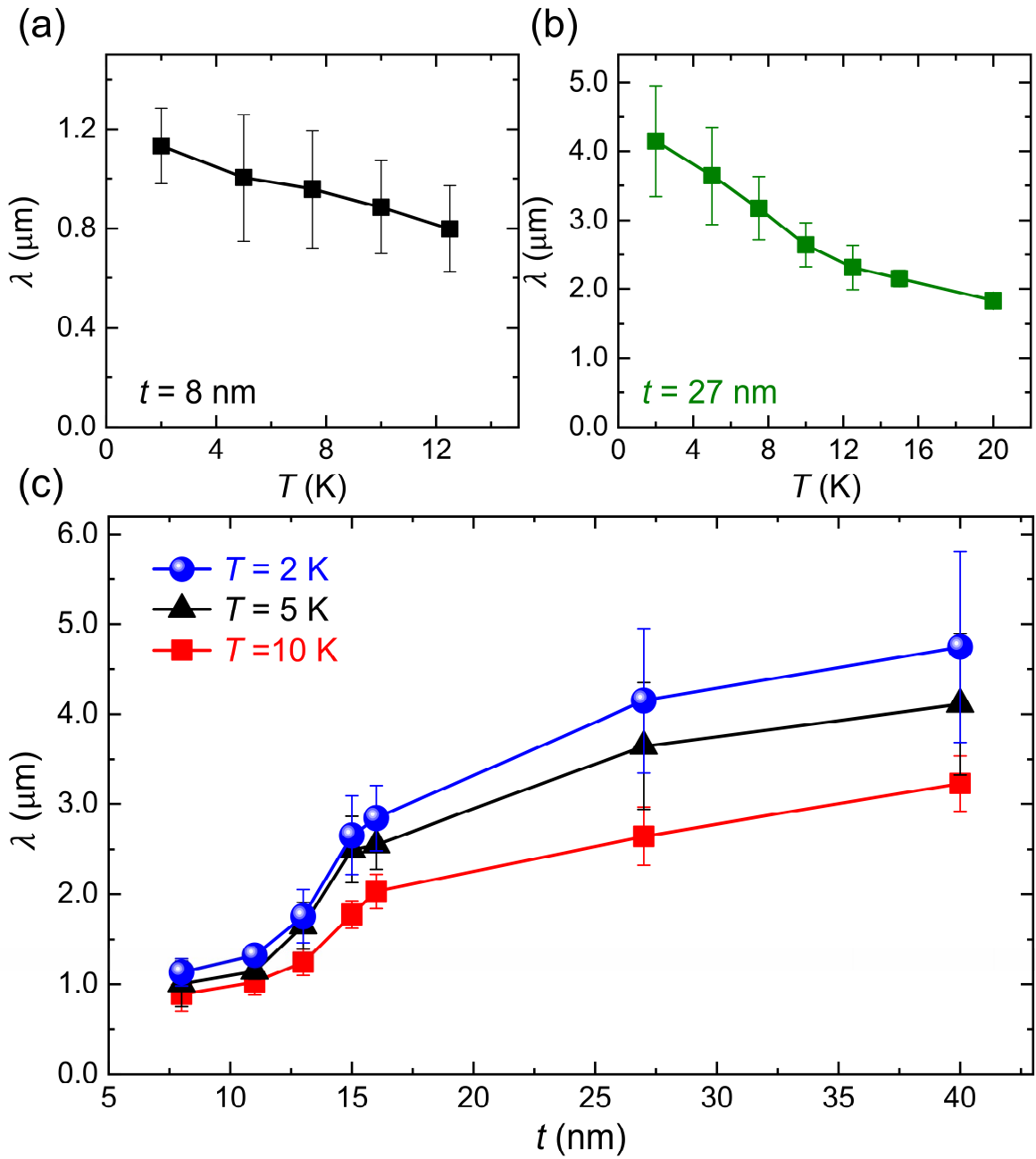

Figure 4 | Thickness dependence of magnon relaxation length of quasi-2D antiferromagnet MnPS$_3$. (a-b) The magnon relaxation length as a function of temperature for the 8 nm and 27 nm quasi-2D MnPS$_3$ devices, respectively. (c) The MnPS$_3$ thickness dependence of the magnon relaxation length obtained at $T = 2$, 5, and 10 K, respectively.

# Supplementary Materials for:

# Magnon transport in quasi-two-dimensional van der Waals antiferromagnets

Wenyu Xing[1,2], Luyi Qiu[1,2], Xirui Wang[1,2], Yunyan Yao[1,2], Yang Ma[1,2], Ranran Cai[1,2], Shuang Jia[1,2,3], X. C. Xie[1,2,3], and Wei Han[1,2*]

[1] International Center for Quantum Materials, School of Physics, Peking University, Beijing 100871, P. R. China

[2] Collaborative Innovation Center of Quantum Matter, Beijing 100871, P. R. China

[3] CAS Center for Excellence in Topological Quantum Computation, University of Chinese Academy of Sciences, Beijing 100190, China

* Correspondence to: weihan@pku.edu.cn

**Figure S1**

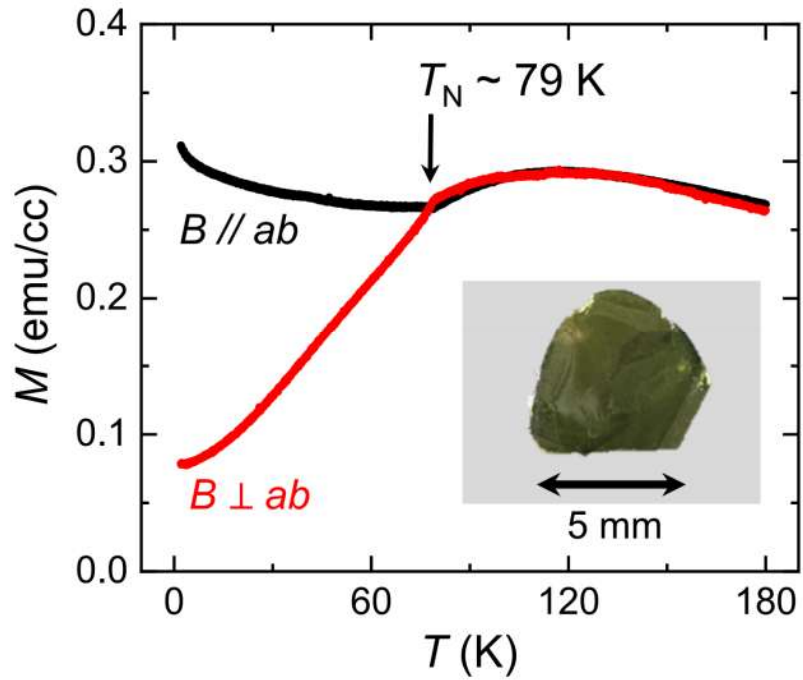

Fig. S1 | Determination of Néel temperature ($T_N$) of bulk MnPS$_3$ single crystal. The temperature dependence of in-plane (black) and out-of-plane (red) magnetization of MnPS$_3$ under the magnetic field of 1 T. Inset: the optical image of a typical MnPS$_3$ single crystal synthesized via the chemical vapor transport method.

**Figure S2**

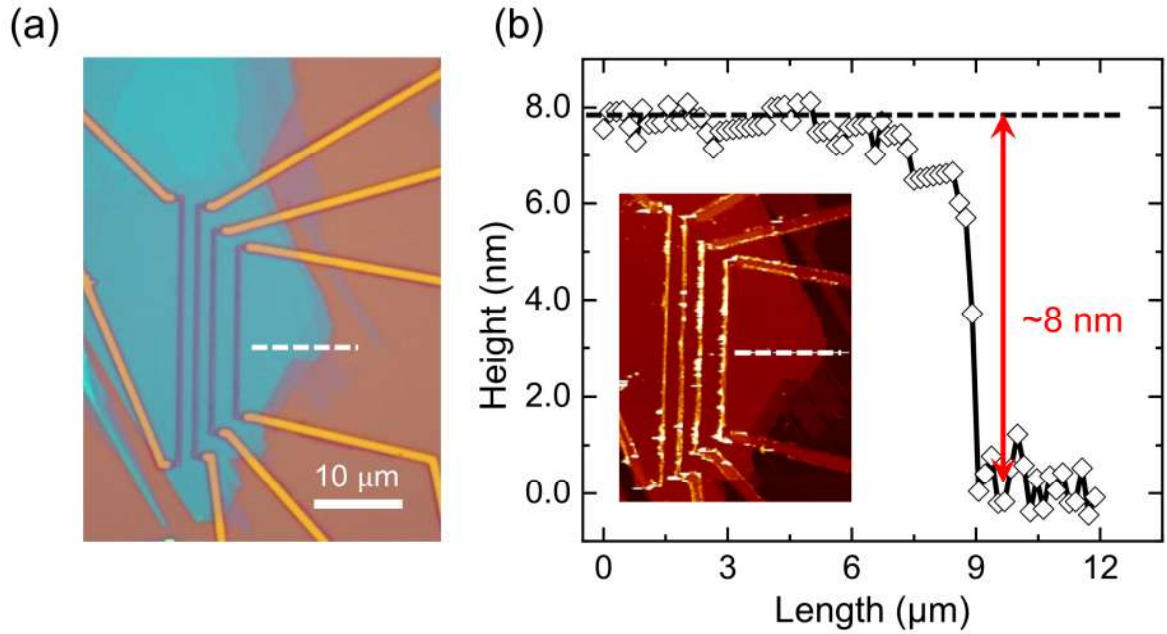

Fig. S2 | Measurement of the quasi-2D MnPS$_3$ thickness via atomic force microscopy. (a) The optical image of the 8 nm MnPS$_3$ device. (b) The height profile across the MnPS$_3$ flake along the white dashed line in a. Inset: the atomic force microscopy image of the 8 nm MnPS$_3$ device.

**Figure S3**

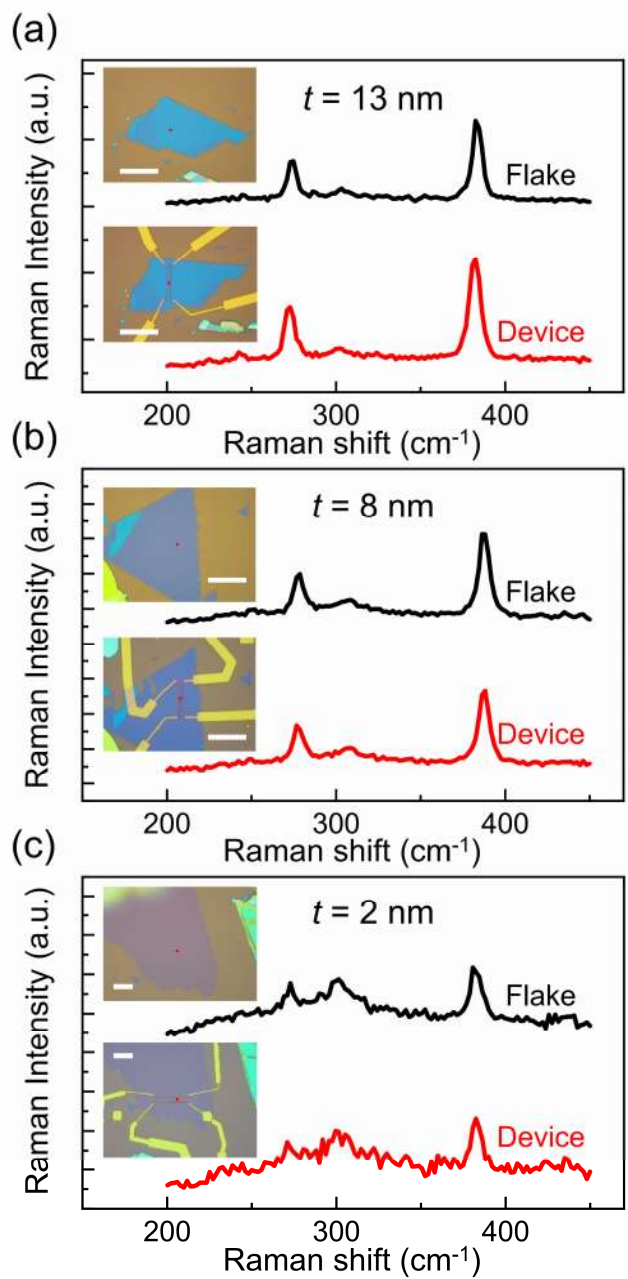

Fig. S3 | Raman characterization of the quasi-2D MnPS$_3$ flakes before and after the device fabrication processes. (a-c) Raman results for three typical MnPS$_3$ samples ($t$ = 13, 8, and 2 nm) before and after the lithography and device fabrication processes. The laser wavelength is 532 nm and power is 0.25 mW. The red dots indicate the position of the laser spots during the Raman measurements. The white lines in optical images of the insets indicate the scale bar of 20 μm.

**Figure S4**

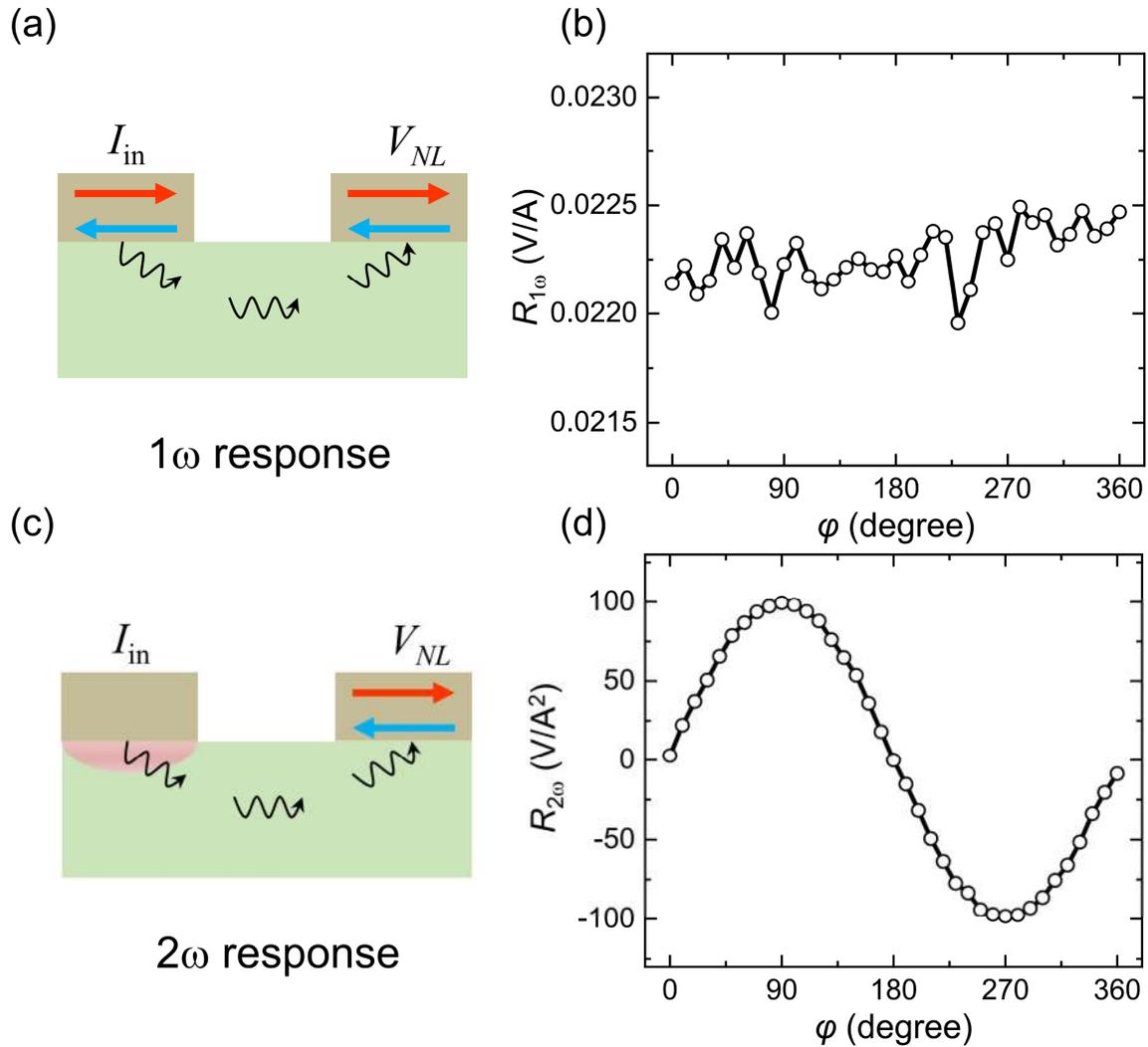

Fig. S4 | The first and second harmonic nonlocal results measured on the 8 nm MnPS$_3$ devices. (a) The illustration of the first harmonic nonlocal measurement, where the magnons are generated via the spin Hall effect of Pt. (b) The first harmonic nonlocal resistance as a function of the magnetic field angle measured at $B = 9$ T and $T = 2$ K. (c) The illustration of the second harmonic nonlocal measurement, where the magnons are generated via thermal means. (d) The second harmonic nonlocal resistance as a function of the magnetic field angle measured at $B = 9$ T and $T = 2$ K.

**Figure S5**

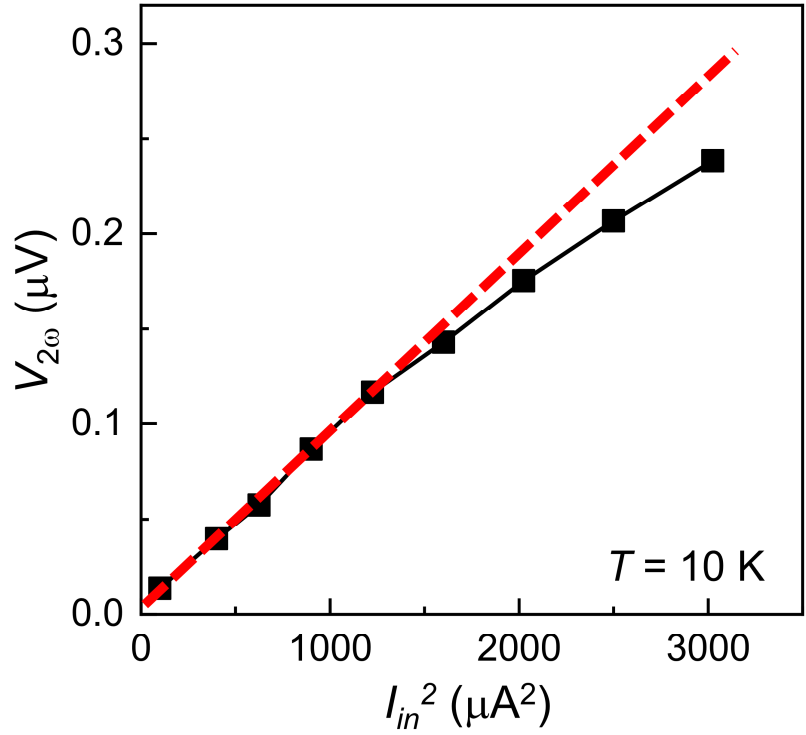

Fig. S5 | The current dependence of the second harmonic voltage measured on the 8 nm MnPS$_3$ device. The measured results (black squares) are obtained at $B = 9$ T and $T = 10$ K. The dashed red line represents the relationship of $V_{2\omega} \sim I_{in}^2$. The deviation from the linear relationship under large current is associated with the increase of the MnPS$_3$ sample temperature.

**Figure S6**

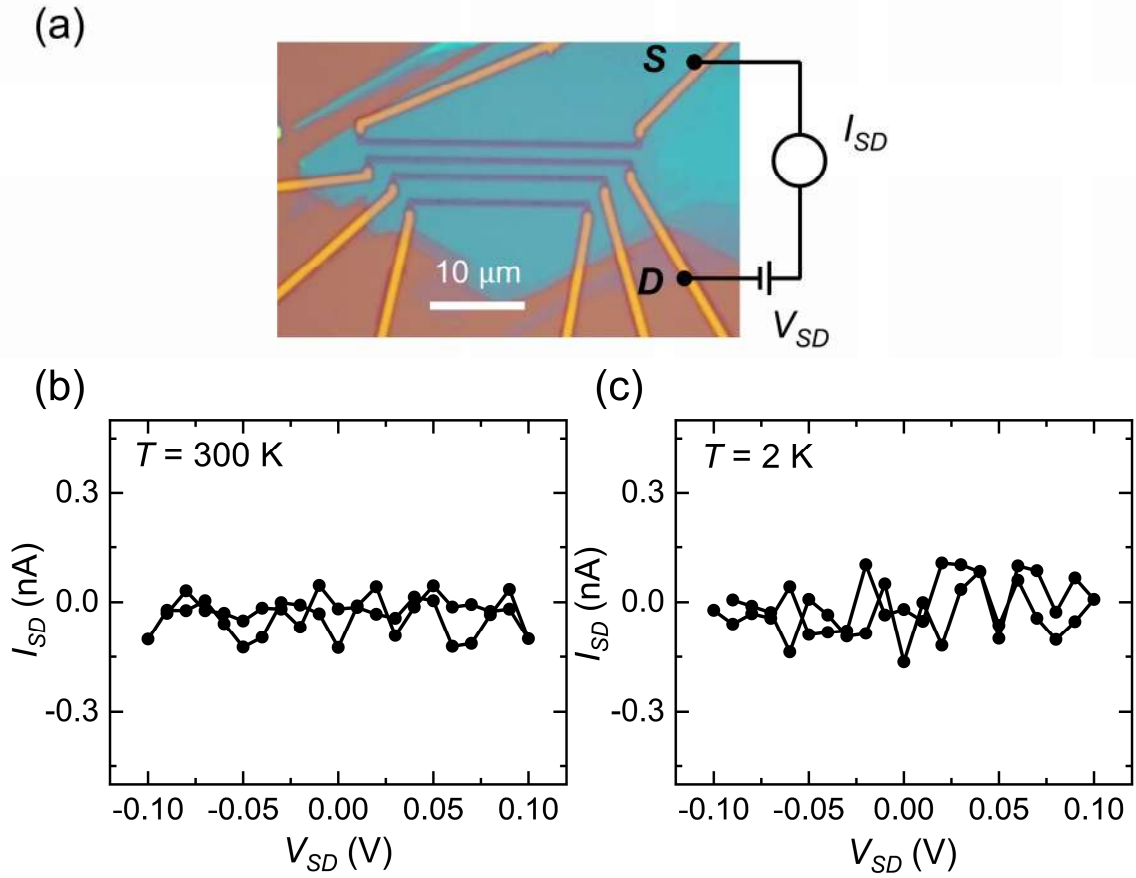

Fig. S6 | The insulating properties of quasi-2D MnPS$_3$ measured on a typical 8 nm flake. (a) Schematic of the two-probe measurement geometry. (b-d) The typical current vs. voltage curves between two Pt electrodes across 8 nm MnPS$_3$ channel at $T$ = 300 K and 2 K, respectively.

**Figure S7**

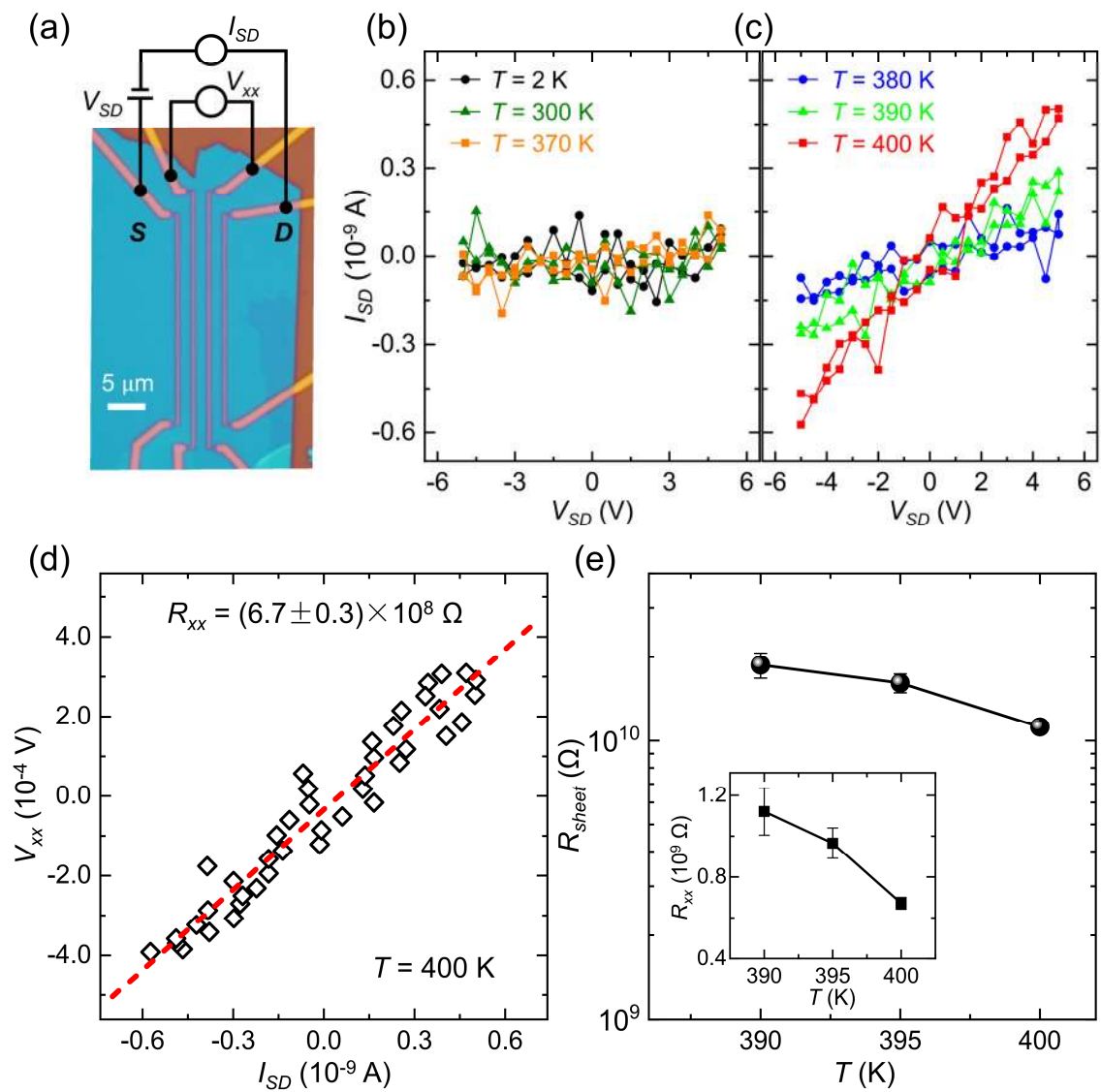

Fig. S7 | The insulating properties of quasi-2D MnPS$_3$ measured on a typical 24 nm flake. (a) Schematic of the two-probe and four-probe measurement geometries on the MnPS$_3$ device. (b-c) The current vs. voltage curves between the source and drain at various temperatures. (e) The $V_{xx}$ vs. current results measured at $T$ = 400 K. The red dashed line represents the best linear fitting curve. (g) Temperature dependence of the sheet resistance of the MnPS$_3$ flake. Inset: temperature dependence of the channel resistance ($R_{xx}$). Below the temperature of 380 K, the MnPS$_3$ is beyond our measurement limit, which further confirms the insulating nature of MnPS$_3$ flakes below 300 K.

**Figure S8**

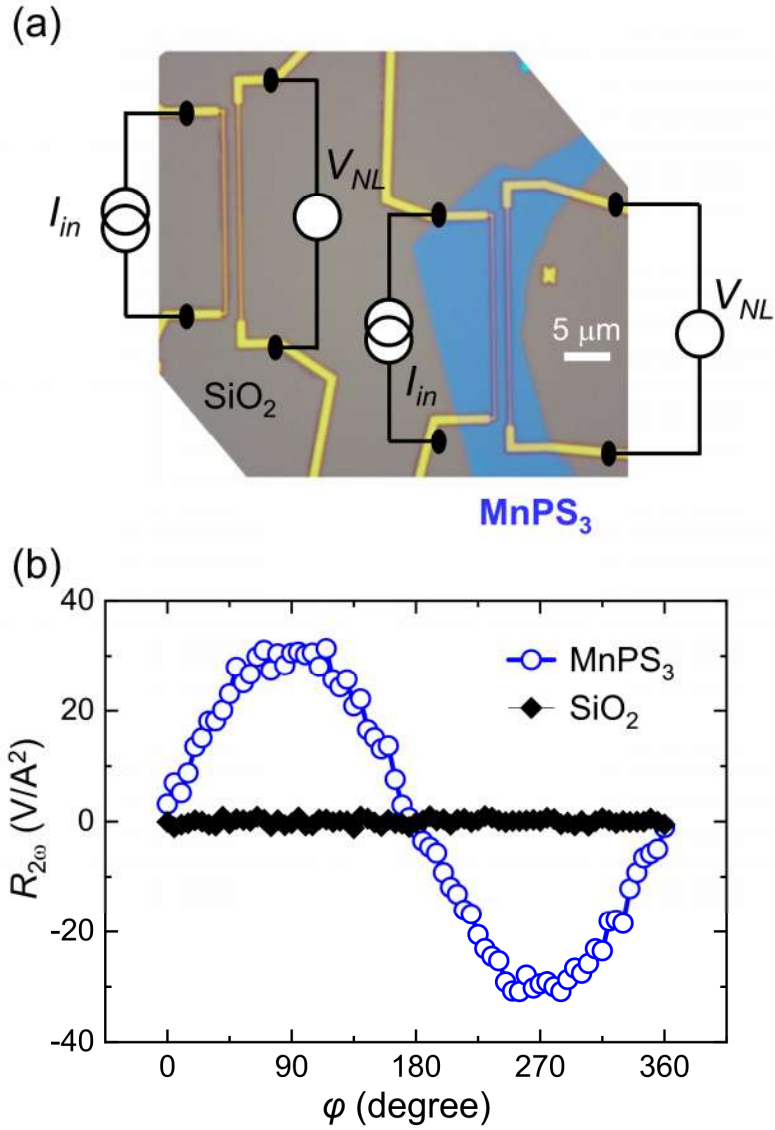

Fig. S8 | Nonlocal measurements on control devices fabricated on $SiO_2$/Si substrates. (a) The optical image of a control device fabricated on $SiO_2$/Si substrate next to a $MnPS_3$ device. (b) The nonlocal resistances measured on the $MnPS_3$ and control devices. The absence of spin signal on the control device rules out the possibility of magnon transport via the $SiO_2$/Si substrate.

**Figure S9**

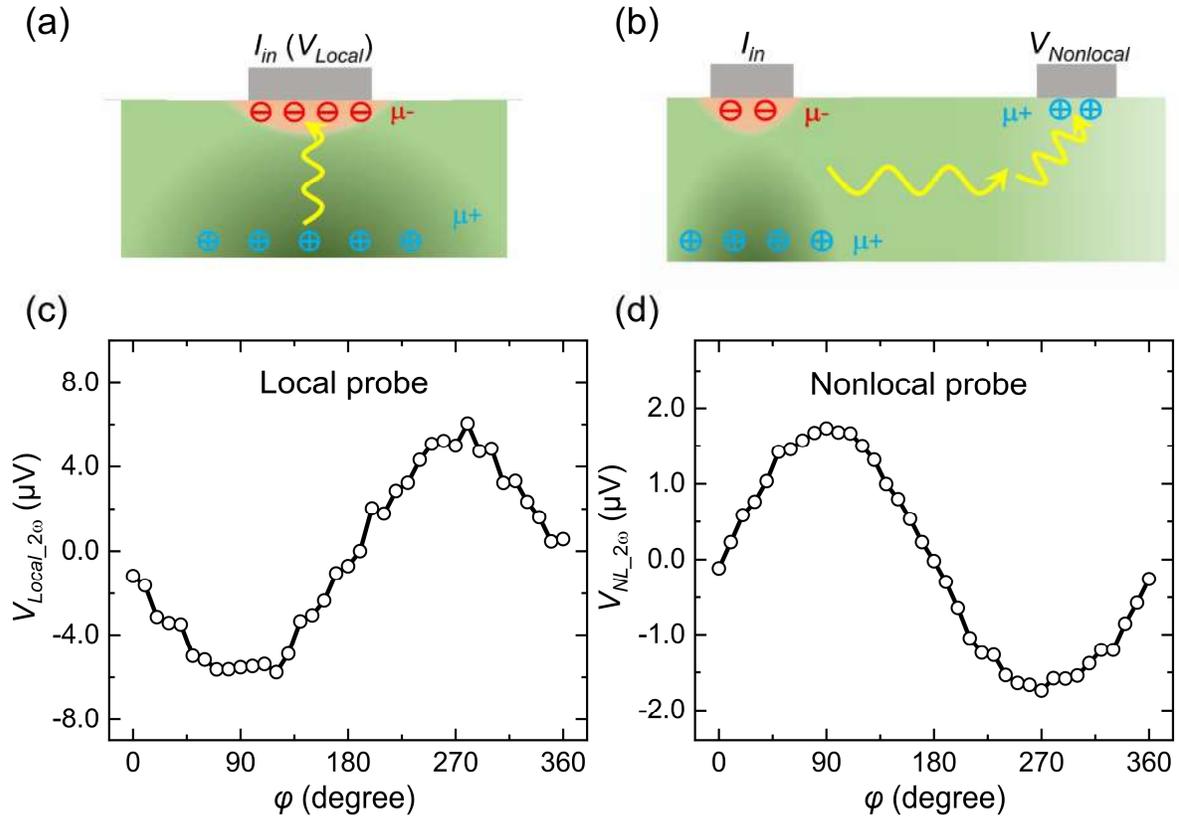

Fig. S9 | Local vs. nonlocal measurements on the MnPS$_3$ magnon device. (a-b) Illustration for the opposite voltage signs measured from the local Pt injector which measures the magnon depletion, and from the nonlocal Pt detector which probes magnon accumulation due to the magnon transport. µ+ and µ– indicate the magnon accumulation and depletion. (c-d) The magnetic field angle dependence of the local voltage measured from the Pt injector and the nonlocal voltage measured from the Pt detector on the same MnPS$_3$ device ($t$ = 13 nm). The spacing between the Pt injector and Pt detector is 2 µm.

**Figure S10**

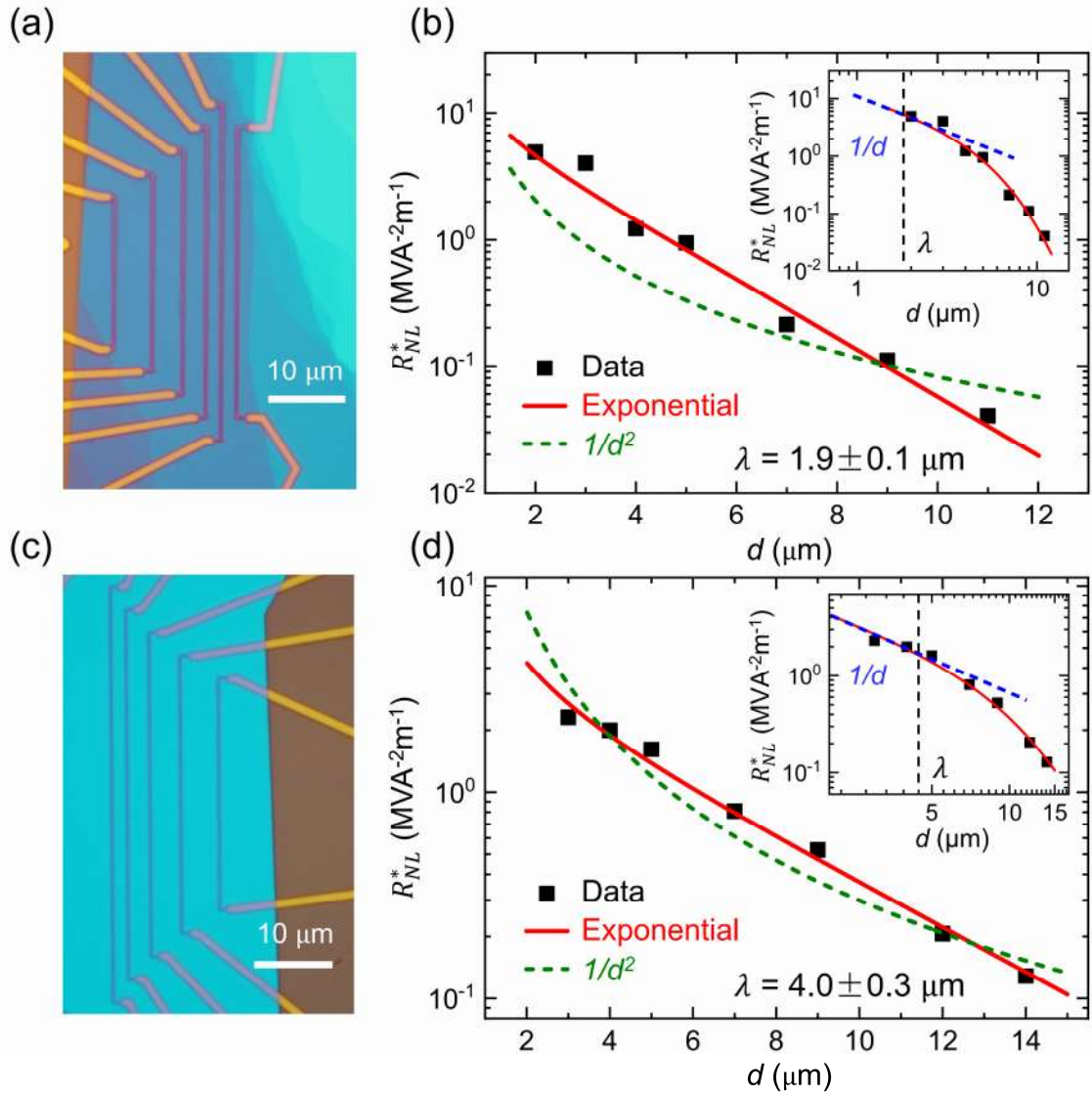

Fig. S10 | Spacing dependence of two extra MnPS$_3$ devices with larger spacings. (a-b) The optical image of the first extra MnPS$_3$ device ($t$ = 13 nm) and the spacing dependence of the normalized nonlocal spin signals ($R^*_{NL}$). No clear spin signal could be obtained for spacings of 12, 14, and 16 μm. (c-d) The optical image of the second extra MnPS$_3$ device ($t$ = 24 nm) and the spacing dependence of $R^*_{NL}$. Clearly, the experimental data (black squares) agree well with the exponential decay (red lines), but not with the *1/d²* decay (green dashed lines). Insets of Figs. (b) and (d) show the similarity between exponential decay (red lines) and *1/d* decay (blue dashed lines) when spacing is smaller than the magnon relaxation length.

**Figure S11**

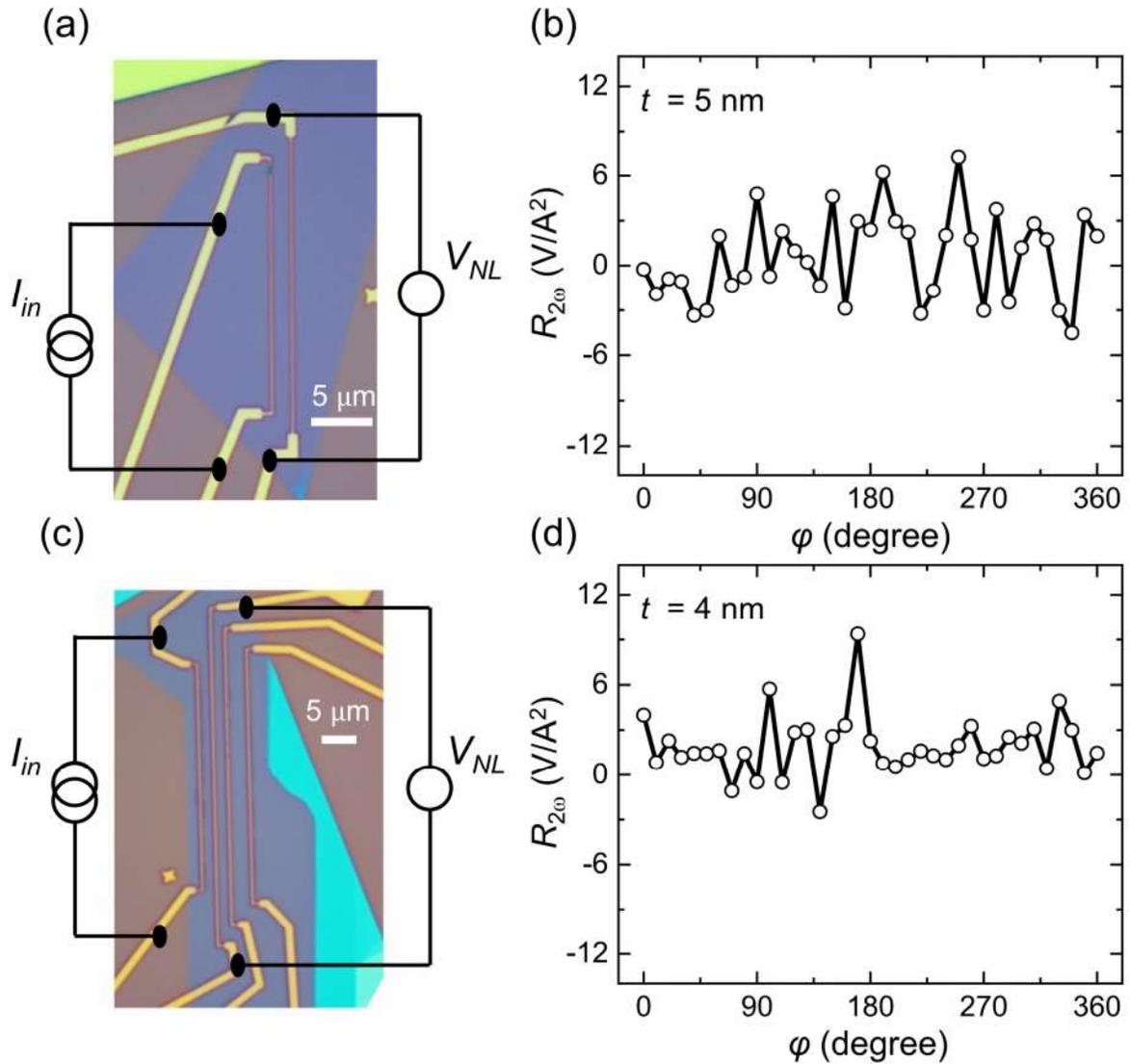

Fig. S11 | The magnon transport measurement on ultrathin 2D MnPS$_3$ devices. (a-b) The optical images and nonlocal magnon transport results of the 5 nm MnPS$_3$ device. (c-d) The optical images and nonlocal magnon transport results of the 4 nm MnPS$_3$ device. The nonlocal magnon transport measurements in (b) and (d) are performed at $B$ = 9 T and $T$ = 2 K.